\documentclass{article}
\usepackage{amssymb}
\usepackage{amsfonts}
\usepackage{amsmath}

\newtheorem{theorem}{Theorem}
\newtheorem{acknowledgement}[theorem]{Acknowledgement}

\input{tcilatex}

\begin{document}

\title{Splitting the Kemmer-Duffin-Petiau Equations}
\author{A. Okni\'{n}ski \\
Physics Division, Politechnika Swietokrzyska, \\
Al. 1000-lecia PP 7, 25-314 Kielce,\\
Poland}
\maketitle

\begin{abstract}
We study internal structure of the Kemmer-Duffin-Petiau equations for spin-$%
0 $ and spin-$1$ mesons. We demonstrate, that the Kemmer-Duffin-Petiau
equations can be splitted into constituent equations, describing particles
with definite mass and broken Lorentz symmetry. We also show that solutions
of the three component constituent equations fulfill the Dirac equation.
\end{abstract}


\section{Introduction}

In recent years there has been a renewed interest\ in the
Kemmer-Duffin-Petiau (KDP) theory describing spin-$0$ and spin-$1$ mesons 
\cite{Kemmer39, Duffin38, Petiau36} due to discovery of a new conserved
four-vector current with positive zeroth component \cite{Ghose}, which can
be thus interpreted as a probability density. A progress was also made in
demonstrating equivalence of the KDP and the Klein-Gordon equations,
especially when interactions are taken into account, c.f. \cite{Lunardi00}
and references therein. The KDP equations has been also studied in the
context of electromagnetic interactions \cite{Beckers95, Now98, Gonen02},
parasupersymmetric quantum mechanics \cite{Beckers95}, EPR type nonlocality 
\cite{Ghose01}, and Riemann-Cartan space-time \cite{Casana02}.

It is well known that the KDP equations contain redundant components - only $%
2\left( 2s+1\right) $\ components are needed to describe free spin-$s$\
particles with nonzero rest masses \cite{Fushchich} while spin-$0$\ and spin-%
$1$\ KDP equations contain $5$\ and $10$\ components, respectively. The
presence of redundant components in KDP equations leads for some
interactions to nonphysical effects such as superluminal velocities \cite%
{Velo69a,Wightman74} (see also Refs. \cite{Johnson61,Wightman68,Velo69b}\
for $s=\frac{3}{2},2$ cases). It is possible however to obtain physically
acceptable equations for arbitrary spin removing redundant components with
use of additional covariant condition \cite{Fushchich}. On the other hand,
presence of redundant components suggests that the KDP equations posses
internal structure. Indeed, a pair of three-component equations, with
solutions fulfilling the five-component spin-$0$\ KDP equation, was found 
\cite{AO}. This internal structure is imperfect in a sense, that although
each of the three-component equations describes a massive particle, its
Lorentz symmetry is broken (i.e. equations are covariant with respect to
boost in one direction and rotation around this axis only). However, these
two three-component equations considered together are Lorentz covariant.

In the present paper we initiate a systematic study of the internal
structure of spin-$0$ and spin-$1$ KDP equations. We shall describe a
systematic procedure of splitting (five-component) spin-$0$\ and
(ten-component)\ spin-$1$\ KDP equations by means of the spinor calculus
into pairs of constituents equations with lesser numbers of components, such
that solutions of the latter equations fulfill the initial KDP equations. In
deep inelastic scattering one is probing the hadron in the infinite momentum
frame. It was first shown by Susskind that the infinite momentum frame is
equivalent to a change of the standard variables $\left(
x^{0},x^{1},x^{2},x^{3}\right) $ into the light-cone variables $\left(
x^{0}+x^{3},x^{0}-x^{3},x^{1},x^{2}\right) $ \cite{Susskind68, Beyer98}.
This suggests that it might be useful to rewrite the tensor KDP equations
within the spinor formalism in which coordinates $\left(
x^{0}+x^{3},x^{0}-x^{3},x^{1}-ix^{2},x^{1}+ix^{2}\right) $, complexifying
the light-cone variables, appear in natural fashion. More exactly, we shall
demonstrate that there is a systematic procedure of splitting
(five-component) spin-$0$ and (ten-component)\ spin-$1$ KDP equations by
means of the spinor calculus into pairs of equations with lesser numbers of
components, such that solutions of the latter equations fulfill the initial
KDP equations. Since mesons are spin-$0$ and spin-$1$ quark-antiquark bound
states it is tempting to recognize the resulting equations as quark
equations but we shall adopt a more cautious approach and will refer to them
as constituent equations. Indeed, we shall show that solutions of
constituent equations fulfill the Dirac equation. Lack of the full
covariance shows that separation of a meson into constituents is frame
dependent (yet is possible, as we shall demonstrate, in arbitrary reference
frame).

The paper is organized as follows. In Section $2$ the Kemmer-Duffin-Petiau
equations for spin $0$ and spin $1$ are described. Elements of spinor
calculus are given in Section $3$ \cite{Corson53, MTW73}. These two Sections
contain also necessary definitions and conventions. In Section $4$ splitting
of the KDP equations into three-component constituent equations is achieved
for $s=0$ (covering in a new way our previous result \cite{AO}). Main
results are described in the next two Sections. In Section $5$ we split the
KDP equations for $s=1$. In Section $6$ we interpret all constituent
equations finding direct relation with the Dirac equation. In the last
Section our results are discussed in the light of several current results
and problems of quark theory.

\section{Kemmer-Duffin-Petiau equations}

In what follows tensor indices are denoted with Greek letters, $\mu=0,1,2,3$%
. We shall use the following convention for the metric tensor: $g^{\mu\nu}=$ 
\textrm{diag}$\left( 1,-1,-1,-1\right) $ and we shall always sum over
repeated indices. Four-momentum operators are defined in natural units ($c=1$%
, $\hslash=1$) as $p^{\mu}=i\frac{\partial}{\partial x_{\mu}}$.

The KDP equations for spin $0$ and $1$ are written as:

\begin{equation}
\beta_{\mu}p^{\mu}\Psi=m\Psi,  \label{KDP-s0,1}
\end{equation}
with $5\times5$ and $10\times10$ matrices $\beta^{\mu}$, respectively, which
fulfill the following commutation relations \cite{Duffin38}: 
\begin{equation}
\beta^{\lambda}\beta^{\mu}\beta^{\nu}+\beta^{\nu}\beta^{\mu}\beta^{\lambda
}=g^{\lambda\mu}\beta^{\nu}+g^{\nu\mu}\beta^{\lambda}.  \label{algebra-b}
\end{equation}

In the case of $5\times5$ (spin-$0$) representation of $\beta^{\mu}$
matrices Eq.(\ref{KDP-s0,1}) is equivalent to the following set of
equations: 
\begin{equation}
\left. 
\begin{array}{ccc}
p^{\mu}\psi & = & m\psi^{\mu} \\ 
p_{\nu}\psi^{\nu} & = & m\psi%
\end{array}
\right\} ,  \label{KDP-s0-1}
\end{equation}
if we define $\Psi$ in (\ref{KDP-s0,1}) as:

\begin{equation}
\Psi=\left( \psi^{\mu},\psi\right) ^{T}=\left( \psi^{0},\psi^{1},\psi
^{2},\psi^{3},\psi\right) ^{T},  \label{wavef-0}
\end{equation}
where $^{T}$ denotes transposition of a matrix. Let us note that Eq.(\ref%
{KDP-s0-1}) can be obtained by factorizing second-order derivatives in the
Klein-Gordon equation $p_{\mu}p^{\mu}\,\psi=m^{2}\psi$.

In the case of $10\times10$ (spin-$1$) representation of matrices $%
\beta^{\mu }$ Eq.(\ref{KDP-s0,1}) reduces to:

\begin{equation}
\left. 
\begin{array}{ccc}
p^{\mu}\psi^{\nu}-p^{\nu}\psi^{\mu} & = & m\psi^{\mu\nu} \\ 
p_{\mu}\psi^{\mu\nu} & = & m\psi^{\nu}%
\end{array}
\right\} ,  \label{KDP-s1-1}
\end{equation}
with the following definition of $\Psi$ in (\ref{KDP-s0,1}): 
\begin{equation}
\Psi=\left( \psi^{\mu\nu},\psi^{\lambda}\right) ^{T}=\left(
\psi^{01},\psi^{02},\psi^{03},\psi^{23},\psi^{31},\psi^{12},\psi^{0},%
\psi^{1},\psi ^{2},\psi^{3}\right) ^{T},  \label{wavef-1}
\end{equation}
where $\psi^{\lambda}$ are real and $\psi^{\mu\nu}$ are purely imaginary (in
alternative formulation we have $-\partial^{\mu}\psi^{\nu}+\partial^{\nu}%
\psi^{\mu}=m\psi^{\mu\nu}$, $\partial_{\mu}\psi^{\mu\nu}=m\psi^{\nu}$, where 
$\psi^{\lambda}$, $\psi^{\mu\nu}$ are real). Because of antisymmetry of $%
\psi^{\mu\nu}$ we have $p_{\nu}\psi^{\nu}=0$ what implies spin $1$
condition. The set of equations (\ref{KDP-s1-1}) was first written by Proca 
\cite{Proca36}.

\section{Elements of spinor calculus}

Two component undotted spinors $\xi_{A}$, where $A$ numbers spinor
components, transform according to representation of the group $SL(2,C)$ , $%
\xi _{A}^{\prime}=\mathbf{S}_{A}^{\ \ B}\xi_{B}$, where $\mathbf{S}\in
SL(2,C)$ is a $2\times2$ complex matrix with unit determinant and $A,B=1,2$.
Analogously, two component dotted spinors $\eta_{\dot{A}}$ transform
according to $\eta_{\dot{A}}^{\prime}=\mathbf{\bar{S}}_{\dot{A}}^{\ \ \dot{B}%
}\eta_{\dot
{B}}$ where $\mathbf{\bar{S}}$ is a matrix complex conjugate to 
$\mathbf{S}$ and $\dot{A},\dot{B}=\dot{1},\dot{2}$. Let us stress again that
we sum over repeated indices. Spinor indices are lowered or raised with help
of metric spinor $\varepsilon_{AB}=\varepsilon_{\dot{A}\dot{B}}=\varepsilon
^{AB}=\varepsilon^{\dot{A}\dot{B}}=\left( 
\begin{array}{cc}
0 & 1 \\ 
-1 & 0%
\end{array}
\right) $. It follows that the scalar product of two spinors, e.g. $\xi
_{A}\zeta^{A}$ is invariant and vanishes automatically for $\xi=\zeta$ since 
$\xi_{A}\zeta^{A}=-\xi^{A}\zeta_{A}$.

A general spinor with $j$ dotted and $k$ undotted indices transforms as $%
\omega_{A_{1}\ldots A_{j}\dot{B}_{1}\ldots\dot{B}_{k}}^{\prime}=\mathbf{S}%
_{A_{1}}^{\ \ C_{1}}\ldots\mathbf{S}_{A_{j}}^{\ \ C_{j}}\mathbf{\bar{S}}_{%
\dot{B}_{1}}^{\ \ \dot{D}_{1}}\ldots\mathbf{\bar{S}}_{\dot{B}_{k}}^{\ \ \dot{%
D}_{k}}\ \omega_{C_{1}\ldots C_{j}\dot{D}_{1}\ldots\dot{D}_{k}}$. Dirac
bispinors are defined as composed from one undotted and one dotted spinor: $%
\Psi=\left( \xi_{1},\xi_{2},\eta_{\dot{1}},\eta _{\dot{2}}\right) ^{T}$ and
thus transform according to reducible representation of $SL(2,C)$ group.

$SL(2,C)$ group generalizes the Lorentz group $SO\left( 1,3\right) $ (more
precisely, $SL(2,C)$ is the simply connected doubly covering group of $%
SO\left( 1,3\right) $) and hence the tensors, i.e. objects transforming
according to the Lorentz group, can be embedded in the spinor algebra. We
shall provide two examples of such embedding, which we shall need later.

Four-vectors $\psi^{\mu}=\left( \psi^{0},\mathbf{\psi}\right) $ and spinors $%
\zeta^{A\dot{B}}$ are related by formula: 
\begin{equation}
\zeta^{A\dot{B}}=\left( \sigma^{0}\psi^{0}+\mathbf{\sigma}\cdot\mathbf{\psi }%
\right) ^{A\dot{B}}=\left( 
\begin{array}{cc}
\zeta^{1\dot{1}} & \zeta^{1\dot{2}} \\ 
\zeta^{2\dot{1}} & \zeta^{2\dot{2}}%
\end{array}
\right) =\left( 
\begin{array}{cc}
\psi^{0}+\psi^{3} & \psi^{1}-i\psi^{2} \\ 
\psi^{1}+i\psi^{2} & \psi^{0}-\psi^{3}%
\end{array}
\right) ,  \label{4vector-spinor}
\end{equation}
where $A,\dot{B}$ number rows and columns, respectively, and $\sigma^{j}$, $%
j=1,2,3$, are the Pauli matrices, $\sigma^{0}$\ is the unit matrix.

Every antisymmetric tensor can be decomposed into selfdual and antiselfdual
parts: $F_{\mu\nu}=-F_{\nu\mu}=F_{\mu\nu}^{S}+F_{\mu\nu}^{A}$, where $\hat
{%
F}_{\mu\nu}^{S}=F_{\mu\nu}^{S}$, $\hat{F}_{\mu\nu}^{A}=-F_{\mu\nu}^{A}$. In
these formulae a definition of a dual tensor was used: $\hat{F}_{\mu\nu }%
\overset{df}{=}\frac{i}{2}\epsilon_{\mu\nu\kappa\lambda}F^{\kappa\lambda}$.
Selfdual and antiselfdual tensors can be expressed by symmetric spinors $%
\xi_{AB}=\xi_{BA}$ and $\eta_{\dot{A}\dot{B}}=\eta_{\dot{B}\dot{A}}$,
respectively. Namely, we have $F_{\mu\nu}^{S}=\Sigma_{\mu\nu AB}\xi^{AB}$, $%
F_{\mu\nu}^{A}=\Sigma_{\mu\nu\dot{A}\dot{B}}\eta^{\dot{A}\dot{B}}$, where $%
\Sigma_{\mu\nu AB}$, $\Sigma_{\mu\nu\dot{A}\dot{B}}$ are appropriate spin
tensors. In explicit form we have: 
\begin{align}
\left( F_{01}^{S},F_{02}^{S},F_{03}^{S}\right) & =i\left(
F_{23}^{S},F_{31}^{S},F_{12}^{S}\right) =\left(
-\xi^{11}+\xi^{22},i\xi^{11}+i\xi^{22},2\xi^{12}\right) ,
\label{SDtensor-spinor} \\
\left( F_{01}^{A},F_{02}^{A},F_{03}^{A}\right) & =-i\left(
F_{23}^{A},F_{31}^{A},F_{12}^{A}\right) =\left( -\eta^{\dot{1}\dot{1}}+\eta
^{\dot{2}\dot{2}},-i\eta^{\dot{1}\dot{1}}-i\eta^{\dot{2}\dot{2}},2\eta ^{%
\dot{1}\dot{2}}\right) .  \label{ASDtensor-spinor}
\end{align}

Spinor calculus abounds in identities. We provide for further convenience
several examples of identities involving spinor $p^{A\dot{B}}$:

\begin{equation}
\begin{array}{ccccccc}
p_{1\dot{1}}p^{1\dot{1}}+p_{2\dot{1}}p^{2\dot{1}} & = & p_{\mu}p^{\mu}, &  & 
p_{1\dot{2}}p^{1\dot{2}}+p_{2\dot{2}}p^{2\dot{2}} & = & p_{\mu}p^{\mu},%
\end{array}
\label{id1}
\end{equation}

\begin{equation}
\begin{array}{ccccccc}
p_{\,\ \dot{B}}^{C}p_{A}^{\ \,\dot{B}} & = & -\delta_{A}^{C}p_{\mu}p^{\mu},
&  & p_{A}^{\,\,\dot{D}}p_{\,\ \dot{B}}^{A} & = & -\delta_{\dot{B}}^{\dot{D}%
}p_{\mu}p^{\mu},%
\end{array}
\label{id2}
\end{equation}

\begin{equation}
\begin{array}{ccccccc}
p_{1}^{\ \,\dot{1}}p_{\,\ \dot{1}}^{2}+p_{1}^{\ \,\dot{2}}p_{\,\ \dot{2}}^{2}
& = & 0, &  & p_{2}^{\ \,\dot{1}}p_{\,\ \dot{1}}^{1}+p_{2}^{\ \,\dot{2}}p_{\
\,\dot{2}}^{1} & = & 0,%
\end{array}
\label{id3}
\end{equation}
which can be verified directly with help of (\ref{4vector-spinor}). For
example, both identities (\ref{id1}) are equivalent to the identity $\left(
p_{0}-p_{3}\right) \left( p_{0}+p_{3}\right) +\left( -p_{1}+ip_{2}\right)
\left( p_{1}+ip_{2}\right) =\left( p_{0}\right) ^{2}-\left( p_{1}\right)
^{2}-\left( p_{2}\right) ^{2}-\left( p_{3}\right) ^{2}$.

\section{Splitting the spin-$0$ Kemmer-Duffin-Petiau equations}

Equations (\ref{KDP-s0-1}) can be written within spinor formalism as:

\begin{equation}
\left. 
\begin{array}{ccc}
p^{A\dot{B}}\psi & = & m\psi^{A\dot{B}} \\ 
p_{A\dot{B}}\psi^{A\dot{B}} & = & 2m\psi%
\end{array}
\right\} .  \label{KDP-s0-2}
\end{equation}

Splitting the last of equations (\ref{KDP-s0-2}), $p_{A\dot{B}}\psi^{A\dot{B}%
}=p_{1\dot{1}}\psi^{1\dot{1}}+p_{2\dot{1}}\psi^{2\dot{1}}+p_{1\dot{2}}\psi^{1%
\dot{2}}+p_{2\dot{2}}\psi^{2\dot{2}}=2m\psi$, we obtain two sets of
equations involving components $\psi^{1\dot{1}},\psi^{2\dot{1}},\psi$ and $%
\psi^{1\dot{2}},\psi^{2\dot{2}},\psi$, respectively:

\begin{equation}
\left. 
\begin{array}{r}
p^{1\dot{1}}\psi =m\psi ^{1\dot{1}} \\ 
p^{2\dot{1}}\psi =m\psi ^{2\dot{1}} \\ 
p_{1\dot{1}}\psi ^{1\dot{1}}+p_{2\dot{1}}\psi ^{2\dot{1}}=m\psi 
\end{array}%
\right\} ,  \label{const-s0-1}
\end{equation}%
\begin{equation}
\left. 
\begin{array}{r}
p^{1\dot{2}}\psi =m\psi ^{1\dot{2}} \\ 
p^{2\dot{2}}\psi =m\psi ^{2\dot{2}} \\ 
p_{1\dot{2}}\psi ^{1\dot{2}}+p_{2\dot{2}}\psi ^{2\dot{2}}=m\psi 
\end{array}%
\right\} ,  \label{const-s0-2}
\end{equation}%
each of which describes particle with mass $m$ (we check this substituting
e.g. $\psi ^{1\dot{1}}$, $\psi ^{2\dot{1}}$ or $\psi ^{1\dot{2}}$, $\psi ^{2%
\dot{2}}$\ into the third equations). The splitting preserving $p_{\mu
}p^{\mu }\psi =m^{2}\psi $ is possible due to spinor identities (\ref{id1}).
Thus solutions of Eqs.(\ref{const-s0-1}), (\ref{const-s0-2}) fulfill the KDP
equations (\ref{KDP-s0-2}). We described these equations in \cite{AO}. From
each of equations(\ref{const-s0-1}), (\ref{const-s0-2}) an identity follows: 
\begin{subequations}
\begin{align}
p^{2\dot{1}}\psi ^{1\dot{1}}& =p^{1\dot{1}}\psi ^{2\dot{1}},
\label{identities0-a} \\
p^{2\dot{2}}\psi ^{1\dot{2}}& =p^{1\dot{2}}\psi ^{2\dot{2}}.
\label{identities0-b}
\end{align}

Equations (\ref{const-s0-1}), (\ref{const-s0-2}) can be written in matrix
form:

\end{subequations}
\begin{equation}
\rho_{\mu}p^{\mu}\Phi=m\Phi,  \label{const-s0-3}
\end{equation}
where $\Phi=\left( \psi^{1\dot{1}},\psi^{2\dot{1}},\psi\right) ^{T}$,

\begin{align}
\rho^{0} & =\left( 
\begin{array}{ccc}
0 & 0 & 1 \\ 
0 & 0 & 0 \\ 
1 & 0 & 0%
\end{array}
\right) ,\rho^{1}=\left( 
\begin{array}{ccc}
0 & 0 & 0 \\ 
0 & 0 & -1 \\ 
0 & 1 & 0%
\end{array}
\right) ,  \label{rho-s0-1a} \\
\rho^{2} & =\left( 
\begin{array}{ccc}
0 & 0 & 0 \\ 
0 & 0 & -i \\ 
0 & -i & 0%
\end{array}
\right) ,\rho^{3}=\left( 
\begin{array}{ccc}
0 & 0 & -1 \\ 
0 & 0 & 0 \\ 
1 & 0 & 0%
\end{array}
\right) ,  \notag
\end{align}
and 
\begin{equation}
\tilde{\rho}_{\mu}p^{\mu}\tilde{\Phi}=m\tilde{\Phi},  \label{const-s0-4}
\end{equation}
where $\tilde{\Phi}=\left( \psi^{1\dot{2}},\psi^{2\dot{2}},\psi\right) ^{T}$,

\begin{align}
\tilde{\rho}^{0} & =\left( 
\begin{array}{ccc}
0 & 0 & 0 \\ 
0 & 0 & 1 \\ 
0 & 1 & 0%
\end{array}
\right) ,\tilde{\rho}^{1}=\left( 
\begin{array}{ccc}
0 & 0 & -1 \\ 
0 & 0 & 0 \\ 
1 & 0 & 0%
\end{array}
\right) ,  \label{rho-s0-1b} \\
\tilde{\rho}^{2} & =\left( 
\begin{array}{ccc}
0 & 0 & i \\ 
0 & 0 & 0 \\ 
i & 0 & 0%
\end{array}
\right) ,\tilde{\rho}^{3}=\left( 
\begin{array}{ccc}
0 & 0 & 0 \\ 
0 & 0 & 1 \\ 
0 & -1 & 0%
\end{array}
\right) .  \notag
\end{align}

Equations (\ref{const-s0-3}), (\ref{const-s0-4}) considered together: 
\begin{equation}
\left( 
\begin{array}{cc}
\rho_{\mu}p^{\mu} & \mathbf{0} \\ 
\mathbf{0} & \tilde{\rho}_{\mu}p^{\mu}%
\end{array}
\right) \left( 
\begin{array}{c}
\Phi \\ 
\tilde{\Phi}%
\end{array}
\right) =m\left( 
\begin{array}{c}
\Phi \\ 
\tilde{\Phi}%
\end{array}
\right) ,  \label{const-s0-5}
\end{equation}
are Lorentz covariant since involve all components of the spinor $\psi ^{A%
\dot{B}}$. Obviously, all solutions of Eq.(\ref{const-s0-5}) satisfy Eq.(\ref%
{KDP-s0-2}) but the reverse is not true.

\section{Splitting the spin-$1$ Kemmer-Duffin-Petiau equations}

KDP equations (\ref{KDP-s1-1}) can be written in spinor form as \cite%
{Corson53}:

\begin{equation}
\left. 
\begin{array}{c}
p_{A}^{\,\,\dot{B}}\zeta _{C\dot{B}}+p_{C}^{\,\,\dot{B}}\zeta _{A\dot{B}%
}=2m\eta _{AC} \\ 
p_{\,\,\dot{B}}^{A}\zeta _{A\dot{D}}+p_{\,\,\dot{D}}^{A}\zeta _{A\dot{B}%
}=2m\chi _{\dot{B}\dot{D}} \\ 
p_{A}^{\,\,\dot{C}}\chi _{\dot{B}\dot{C}}+p_{\,\,\dot{B}}^{C}\eta
_{AC}=-2m\zeta _{A\dot{B}}%
\end{array}%
\right\} .  \label{KDP-s1-2}
\end{equation}%
It is possible to split the spinor form of the KDP equations (\ref{KDP-s1-2}%
) to get two equations for spinors $\chi _{\dot{B}\dot{D}}$, $\zeta _{A\dot{B%
}}$ and $\eta _{AC}$, $\zeta _{A\dot{B}}$: 
\begin{equation}
\left. 
\begin{array}{l}
p_{A}^{\ \,\dot{B}}\zeta _{C\dot{B}}=m\eta _{AC},\quad \eta _{AC}=\eta _{CA}
\\ 
p_{\,\ \dot{B}}^{C}\eta _{AC}=-m\zeta _{A\dot{B}}%
\end{array}%
\right\} ,  \label{KDP-s1-3a}
\end{equation}%
\begin{equation}
\left. 
\begin{array}{l}
p_{\,\ \dot{B}}^{A}\zeta _{A\dot{D}}=m\chi _{\dot{B}\dot{D}},\quad \chi _{%
\dot{B}\dot{D}}=\chi _{\dot{D}\dot{B}} \\ 
p_{A}^{\,\,\dot{D}}\chi _{\dot{B}\dot{D}}=-m\zeta _{A\dot{B}}%
\end{array}%
\right\} ,  \label{KDP-s1-3b}
\end{equation}%
respectively. The splitting possible due to spinor identities (\ref{id2}).
Thus solutions of Eqs.(\ref{KDP-s1-3a}), (\ref{KDP-s1-3b}) fulfill the KDP
equations (\ref{KDP-s1-2}).

The spinor equations (\ref{KDP-s1-3a}), (\ref{KDP-s1-3b}) describe spin-$1$
bosons \cite{Lopusz78} where spinors $\eta_{CA}$, $\chi_{\dot{D}\dot{B}}$
correspond to selfdual or antiselfdual antisymmetric tensors $\psi^{\mu\nu}$%
, respectively. Each of the above equations is covariant except from space
reflection but both equations taken together are fully covariant. These
equations written in tensor form, $\beta^{\mu}p_{\mu}\Psi=m\Psi$, $\Psi=%
\left[ \psi_{01},\psi_{02},\psi_{03},\psi_{0},\psi_{1},\psi_{2},\psi_{3}%
\right] ^{T}$ where $\psi^{\mu\nu}$ are selfdual or antiselfdual
antisymmetric tensors, with $7\times7$ matrices $\beta^{\mu}$ fulfilling Eq.(%
\ref{Tzou}), are the Hagen-Hurley equations \cite{HagHur, Beckers95}.

We shall now split the spinor form of the Hagen-Hurley equations (\ref%
{KDP-s1-3a}), (\ref{KDP-s1-3b}) to arrive at the main result of this
Section. To this end equation (\ref{KDP-s1-3a}) is written in explicit form.
Since $\eta_{12}=\eta_{21}\equiv\eta$, the left hand side of the subequation 
$p_{1}^{\ \,\dot{B}}\zeta_{2\dot{B}}=\eta_{12}$ is decomposed into symmetric
and antisymmetric parts, the latter equal zero:

\begin{subequations}
\begin{align}
p_{1}^{\ \,\dot{1}}\zeta_{1\dot{1}}+p_{1}^{\ \,\dot{2}}\zeta_{1\dot{2}} &
=m\eta_{11},  \label{KDP-s1-a} \\
\tfrac{1}{2}\left( p_{1}^{\ \,\dot{1}}\zeta_{2\dot{1}}+p_{1}^{\ \,\dot{2}%
}\zeta_{2\dot{2}}+p_{2}^{\ \,\dot{1}}\zeta_{1\dot{1}}+p_{2}^{\ \,\dot{2}%
}\zeta_{1\dot{2}}\right) & =m\eta,  \label{KDP-s1-b} \\
p_{1}^{\ \,\dot{1}}\zeta_{2\dot{1}}+p_{1}^{\ \,\dot{2}}\zeta_{2\dot{2}%
}-p_{2}^{\ \,\dot{1}}\zeta_{1\dot{1}}-p_{2}^{\ \,\dot{2}}\zeta_{1\dot{2}} &
=0,  \label{KDP-s1-c} \\
p_{2}^{\ \,\dot{1}}\zeta_{2\dot{1}}+p_{2}^{\ \,\dot{2}}\zeta_{2\dot{2}} &
=m\eta_{22},  \label{KDP-s1-d} \\
&  \notag \\
p_{\,\ \dot{1}}^{1}\eta_{11}+p_{\,\ \dot{1}}^{2}\eta & =-m\zeta_{1\dot{1}},
\label{KDP-s1-e} \\
p_{\,\ \dot{2}}^{1}\eta_{11}+p_{\,\ \dot{2}}^{2}\eta & =-m\zeta_{1\dot{2}},
\label{KDP-s1-f} \\
p_{\,\ \dot{1}}^{1}\eta+p_{\,\ \dot{1}}^{2}\eta_{22} & =-m\zeta_{2\dot{1}},
\label{KDP-s1-g} \\
p_{\,\ \dot{2}}^{1}\eta+p_{\,\ \dot{2}}^{2}\eta_{22} & =-m\zeta_{2\dot{2}}
\label{KDP-s1-h}
\end{align}
Let us note that vanishing of the antisymmetric part, i.e. the third
equation, expresses spin-$1$ condition $p_{\mu}\psi^{\mu}=0$, what can be
verified directly with help of Eq.(\ref{4vector-spinor}).

We shall demonstrate that the set of eight equations above is equivalent to
the following equations:

\end{subequations}
\begin{subequations}
\begin{align}
p_{\,\,\dot{1}}^{1}\eta_{11} & =-m\hat{\zeta}_{1\dot{1}},  \label{const-s1-a}
\\
p_{\,\,\dot{2}}^{1}\eta_{11} & =-m\hat{\zeta}_{1\dot{2}},  \label{const-s1-b}
\\
p_{1}^{\ \,\dot{1}}\hat{\zeta}_{1\dot{1}}+p_{1}^{\ \,\dot{2}}\hat{\zeta }_{1%
\dot{2}} & =m\eta_{11},  \label{const-s1-c}
\end{align}

\end{subequations}
\begin{subequations}
\begin{align}
p_{\,\,\dot{1}}^{2}\eta_{22} & =-m\hat{\zeta}_{2\dot{1}},  \label{const-s1-d}
\\
p_{\,\,\dot{2}}^{2}\eta_{22} & =-m\hat{\zeta}_{2\dot{2}},  \label{const-s1-e}
\\
p_{2}^{\,\,\dot{1}}\hat{\zeta}_{2\dot{1}}+p_{2}^{\,\,\dot{2}}\hat{\zeta }_{2%
\dot{2}} & =m\eta_{22},  \label{const-s1-f}
\end{align}

\end{subequations}
\begin{equation}
p_{\mu}p^{\mu}\eta=m^{2}\eta,  \label{const-s1-g}
\end{equation}
i.e. can be splitted into Eqs.(\ref{const-s1-a}, \ref{const-s1-b}, \ref%
{const-s1-c}), Eqs.(\ref{const-s1-d}, \ref{const-s1-e}, \ref{const-s1-f}),
and Eq.(\ref{const-s1-g}), where $\hat{\zeta}_{1\dot{1}}\overset{df}{=}%
\zeta_{1\dot{1}}+\frac{p_{\,\ \dot{1}}^{2}}{m}\eta$, $\hat{\zeta}_{1\dot{2}}%
\overset{df}{=}\zeta_{1\dot{2}}+\frac{p_{\,\ \dot{2}}^{2}}{m}\eta$, $\hat{%
\zeta}_{2\dot{1}}\overset{df}{=}\zeta_{2\dot{1}}+\frac{p_{\,\ \,\dot{1}}^{1}%
}{m}\eta$, $\hat{\zeta}_{2\dot{2}}\overset{df}{=}\zeta_{2\dot{2}}+\frac{%
p_{\,\ \dot{2}}^{1}}{m}\eta$.

Indeed, $p_{1}^{\ \,\dot{1}}\hat{\zeta}_{1\dot{1}}+p_{1}^{\ \,\dot{2}}\hat{%
\zeta}_{1\dot{2}}=p_{1}^{\ \,\dot{1}}\zeta_{1\dot{1}}+p_{1}^{\ \,\dot{2}%
}\zeta_{1\dot{2}}$ and $p_{2}^{\ \,\dot{1}}\hat{\zeta}_{2\dot{1}}+p_{2}^{\ \,%
\dot{2}}\hat{\zeta}_{2\dot{2}}=p_{2}^{\ \,\dot{1}}\zeta_{2\dot{1}}+p_{2}^{\
\,\dot{2}}\zeta_{2\dot{2}}$ due to spinor identities (\ref{id3}). Hence Eqs.(%
\ref{const-s1-a}), (\ref{const-s1-b}), (\ref{const-s1-c}) are identical with
equations (\ref{KDP-s1-e}), (\ref{KDP-s1-f}), (\ref{KDP-s1-a}) and Eqs.(\ref%
{const-s1-d}), (\ref{const-s1-e}), (\ref{const-s1-f}) are identical with
equations (\ref{KDP-s1-g}), (\ref{KDP-s1-h}), (\ref{KDP-s1-d}). Furthermore,
from Eqs.(\ref{const-s1-a}), (\ref{const-s1-b}) and Eqs.(\ref{const-s1-d}), (%
\ref{const-s1-e}) two identities follow: 
\begin{subequations}
\begin{align}
p_{\,\,\dot{2}}^{1}\hat{\zeta}_{1\dot{1}} & =p_{\,\,\dot{1}}^{1}\hat{\zeta }%
_{1\dot{2}},  \label{identities1-a} \\
p_{\,\,\dot{2}}^{2}\hat{\zeta}_{2\dot{1}} & =p_{\,\,\dot{1}}^{2}\hat{\zeta }%
_{2\dot{2}}.  \label{identities1-b}
\end{align}
Now, $\left( p_{\,\ \dot{2}}^{1}\hat{\zeta}_{1\dot{1}}-p_{\,\ \dot{1}}^{1}%
\hat{\zeta}_{1\dot{2}}\right) +\left( p_{\,\ \dot{2}}^{2}\hat{\zeta }_{2\dot{%
1}}-p_{\,\ \dot{1}}^{2}\hat{\zeta}_{2\dot{2}}\right) =0$ reduces to $p_{\,\ 
\dot{2}}^{1}\zeta_{1\dot{1}}-p_{\,\ \dot{1}}^{1}\zeta_{1\dot{2}}+p_{\,\ \dot{%
2}}^{2}\zeta_{2\dot{1}}-p_{\,\ \dot{1}}^{2}\zeta_{2\dot{2}}=0$ and implies
spin-$1$ condition $p_{\mu}\psi^{\mu}=0$, where $\zeta^{A\dot{B}}=\left(
\sigma^{0}\psi^{0}+\mathbf{\sigma}\cdot\mathbf{\psi}\right) ^{A\dot{B}}$ -
this condition is thus equivalent to Eq.(\ref{KDP-s1-c}). On the other hand,
it can be directly verified using definition of spinor $\hat{\zeta}_{A\dot{B}%
}$ and rearranging indices of the spinor $p_{\ \ \dot
{B}}^{A}$ to get $%
p_{C}^{\ \ \dot{D}}$, that $\left( p_{\,\ \dot{2}}^{1}\hat{\zeta}_{1\dot{1}%
}-p_{\,\ \dot{1}}^{1}\hat{\zeta}_{1\dot{2}}\right) -\left( p_{\,\ \dot{2}%
}^{2}\hat{\zeta}_{2\dot{1}}-p_{\,\ \dot{1}}^{2}\hat{\zeta}_{2\dot{2}}\right)
=0$ is equivalent to equality $p_{1}^{\ \,\dot{1}}\zeta_{2\dot{1}}+p_{1}^{\
\,\dot{2}}\zeta_{2\dot{2}}+p_{2}^{\ \,\dot{1}}\zeta_{1\dot{1}}+p_{2}^{\ \,%
\dot{2}}\zeta_{1\dot{2}}=2\frac{p_{\mu}p^{\mu}}{m}\eta$ and becomes the
equation (\ref{KDP-s1-b}) due to (\ref{const-s1-g}).

The three component equations can be written in matrix form as $\rho_{\mu
}p^{\mu}\Psi=m\Psi$:

\end{subequations}
\begin{align}
\rho^{0} & =\left( 
\begin{array}{ccc}
0 & 0 & 0 \\ 
0 & 0 & -1 \\ 
0 & -1 & 0%
\end{array}
\right) ,\rho^{1}=\left( 
\begin{array}{ccc}
0 & 0 & -1 \\ 
0 & 0 & 0 \\ 
1 & 0 & 0%
\end{array}
\right) ,  \label{rho-s1-1a} \\
\rho^{2} & =\left( 
\begin{array}{ccc}
0 & 0 & i \\ 
0 & 0 & 0 \\ 
i & 0 & 0%
\end{array}
\right) ,\rho^{3}=\left( 
\begin{array}{ccc}
0 & 0 & 0 \\ 
0 & 0 & 1 \\ 
0 & -1 & 0%
\end{array}
\right) ,  \notag
\end{align}
$\Psi=\left( \hat{\zeta}_{1\dot{1}},\hat{\zeta}_{1\dot{2}},\eta_{11}\right)
^{T}$ and $\tilde{\rho}_{\mu}p^{\mu}\tilde{\Psi}=m\tilde{\Psi}$:

\begin{align}
\tilde{\rho}^{0} & =\left( 
\begin{array}{ccc}
0 & 0 & 1 \\ 
0 & 0 & 0 \\ 
1 & 0 & 0%
\end{array}
\right) ,\tilde{\rho}^{1}=\left( 
\begin{array}{ccc}
0 & 0 & 0 \\ 
0 & 0 & 1 \\ 
0 & -1 & 0%
\end{array}
\right) ,  \label{rho-s1-1b} \\
\tilde{\rho}^{2} & =\left( 
\begin{array}{ccc}
0 & 0 & 0 \\ 
0 & 0 & i \\ 
0 & i & 0%
\end{array}
\right) ,\tilde{\rho}^{3}=\left( 
\begin{array}{ccc}
0 & 0 & 1 \\ 
0 & 0 & 0 \\ 
-1 & 0 & 0%
\end{array}
\right) ,  \notag
\end{align}
$\tilde{\Psi}=\left( \hat{\zeta}_{2\dot{1}},\hat{\zeta}_{2\dot{2}},\eta
_{22}\right) ^{T}$.

Analogously, we split equations (\ref{KDP-s1-3b}): 
\begin{equation}
\left. 
\begin{tabular}{rrr}
$p_{\,\ \dot{1}}^{1}\zeta _{1\dot{1}}+p_{\,\ \dot{1}}^{2}\zeta _{2\dot{1}}$
& $=$ & $m\chi _{\dot{1}\dot{1}}$ \\ 
$\tfrac{1}{2}\left( p_{\,\ \dot{1}}^{1}\zeta _{1\dot{2}}+p_{\,\ \dot{1}%
}^{2}\zeta _{2\dot{2}}+p_{\,\ \dot{2}}^{1}\zeta _{1\dot{1}}+p_{\,\ \dot{2}%
}^{2}\zeta _{2\dot{1}}\right) $ & $=$ & $m\chi \ \ $ \\ 
$p_{\,\ \dot{1}}^{1}\zeta _{1\dot{2}}+p_{\,\ \dot{1}}^{2}\zeta _{2\dot{2}%
}-p_{\,\ \dot{2}}^{1}\zeta _{1\dot{1}}-p_{\,\ \dot{2}}^{2}\zeta _{2\dot{1}}$
& $=$ & $0\quad \ \ $ \\ 
$p_{\,\ \dot{2}}^{1}\zeta _{1\dot{2}}+p_{\,\ \dot{2}}^{2}\zeta _{2\dot{2}}$
& $=$ & $m\chi _{\dot{2}\dot{2}}$ \\ 
$p_{1}^{\,\,\dot{1}}\chi _{\dot{1}\dot{1}}+p_{1}^{\,\,\dot{2}}\chi $ & $=$ & 
$-m\zeta _{1\dot{1}}$ \\ 
$p_{2}^{\,\,\dot{1}}\chi _{\dot{1}\dot{1}}+p_{2}^{\,\,\dot{2}}\chi $ & $=$ & 
$-m\zeta _{2\dot{1}}$ \\ 
$p_{1}^{\,\,\dot{1}}\chi +p_{1}^{\,\,\dot{2}}\chi _{\dot{2}\dot{2}}$ & $=$ & 
$-m\zeta _{1\dot{2}}$ \\ 
$p_{2}^{\,\,\dot{1}}\chi +p_{2}^{\,\,\dot{2}}\chi _{\dot{2}\dot{2}}$ & $=$ & 
$-m\zeta _{2\dot{2}}$%
\end{tabular}%
\ \right\} ,  \label{KDP-s1-3b1}
\end{equation}%
where $\chi _{\dot{1}\dot{2}}=\chi _{\dot{2}\dot{1}}\equiv \chi $ (again the
third equation is equivalent to spin-$1$ condition $p_{\mu }\psi ^{\mu }=0$%
). We thus get:

\begin{equation}
\left. 
\begin{array}{r}
p_{1}^{\,\,\dot{1}}\chi_{\dot{1}\dot{1}}=-m\check{\zeta}_{1\dot{1}} \\ 
p_{2}^{\,\,\dot{1}}\chi_{\dot{1}\dot{1}}=-m\check{\zeta}_{2\dot{1}} \\ 
p_{\,\ \dot{1}}^{1}\check{\zeta}_{1\dot{1}}+p_{\,\ \dot{1}}^{2}\check{\zeta }%
_{2\dot{1}}=m\chi_{\dot{1}\dot{1}}%
\end{array}
\right\} ,  \label{const-s1-2a}
\end{equation}%
\begin{equation}
\left. 
\begin{array}{r}
p_{1}^{\,\,\dot{2}}\chi_{\dot{2}\dot{2}}=-m\check{\zeta}_{1\dot{2}} \\ 
p_{2}^{\,\,\dot{2}}\chi_{\dot{2}\dot{2}}=-m\check{\zeta}_{2\dot{2}} \\ 
p_{\,\ \dot{2}}^{1}\check{\zeta}_{1\dot{2}}+p_{\,\ \dot{2}}^{2}\check{\zeta }%
_{2\dot{2}}=m\chi_{\dot{2}\dot{2}}%
\end{array}
\right\} ,  \label{const-s1-2b}
\end{equation}%
\begin{equation}
p_{\mu}p^{\mu}\chi=m^{2}\chi,  \label{const-s1-2c}
\end{equation}
where $\check{\zeta}_{1\dot{1}}\overset{df}{=}\zeta_{1\dot{1}}+\frac {%
p_{1}^{\,\,\dot{2}}}{m}\chi$, $\check{\zeta}_{2\dot{1}}\overset{df}{=}%
\zeta_{2\dot{1}}+\frac{p_{2}^{\,\,\dot{2}}}{m}\chi$, $\check{\zeta}_{1\dot{2}%
}\overset{df}{=}\zeta_{1\dot{2}}+\frac{p_{1}^{\,\,\dot{1}}}{m}\chi$, $\check{%
\zeta}_{2\dot{2}}\overset{df}{=}\zeta_{2\dot{2}}+\frac{p_{2}^{\,\,\dot{1}}}{m%
}\chi$. Equations (\ref{const-s1-2a}), (\ref{const-s1-2b}), (\ref%
{const-s1-2c}) with these definitions are equivalent to (\ref{KDP-s1-3b1})
due to appropriate spinor identities as well as to identities: 
\begin{subequations}
\begin{align}
p_{2}^{\,\,\dot{1}}\check{\zeta}_{1\dot{1}} & =p_{1}^{\,\,\dot{1}}\check{%
\zeta}_{2\dot{1}},  \label{identities1-c} \\
p_{2}^{\,\,\dot{2}}\check{\zeta}_{1\dot{2}} & =p_{1}^{\,\,\dot{2}}\check{%
\zeta}_{2\dot{2}},  \label{identities1-d}
\end{align}
which follow from (\ref{const-s1-2a}), (\ref{const-s1-2b}).

The equations (\ref{const-s1-2a}), (\ref{const-s1-2b}) can be written in
matrix form $\rho_{\mu}p^{\mu}\Psi=m\Psi$:

\end{subequations}
\begin{align}
\rho^{0} & =\left( 
\begin{array}{ccc}
0 & 0 & 0 \\ 
0 & 0 & -1 \\ 
0 & -1 & 0%
\end{array}
\right) ,\rho^{1}=\left( 
\begin{array}{ccc}
0 & 0 & -1 \\ 
0 & 0 & 0 \\ 
1 & 0 & 0%
\end{array}
\right) ,  \label{rho-s1-2a} \\
\rho^{2} & =\left( 
\begin{array}{ccc}
0 & 0 & -i \\ 
0 & 0 & 0 \\ 
-i & 0 & 0%
\end{array}
\right) ,\rho^{3}=\left( 
\begin{array}{ccc}
0 & 0 & 0 \\ 
0 & 0 & 1 \\ 
0 & -1 & 0%
\end{array}
\right) ,  \notag
\end{align}
$\Psi=\left( \check{\zeta}_{1\dot{1}},\check{\zeta}_{2\dot{1}},\chi_{\dot
{1%
}\dot{1}}\right) ^{T}$ and $\tilde{\rho}_{\mu}p^{\mu}\tilde{\Psi}=m\tilde{%
\Psi}$:

\begin{align}
\tilde{\rho}^{0} & =\left( 
\begin{array}{ccc}
0 & 0 & 1 \\ 
0 & 0 & 0 \\ 
1 & 0 & 0%
\end{array}
\right) ,\tilde{\rho}^{1}=\left( 
\begin{array}{ccc}
0 & 0 & 0 \\ 
0 & 0 & 1 \\ 
0 & -1 & 0%
\end{array}
\right) ,  \label{rho-s1-2b} \\
\tilde{\rho}^{2} & =\left( 
\begin{array}{ccc}
0 & 0 & 0 \\ 
0 & 0 & -i \\ 
0 & -i & 0%
\end{array}
\right) ,\tilde{\rho}^{3}=\left( 
\begin{array}{ccc}
0 & 0 & 1 \\ 
0 & 0 & 0 \\ 
-1 & 0 & 0%
\end{array}
\right) ,  \notag
\end{align}
$\tilde{\Psi}=\left( \check{\zeta}_{1\dot{2}},\check{\zeta}_{2\dot{2}},\chi_{%
\dot{2}\dot{2}}\right) ^{T}$. Let us note that matrices (\ref{rho-s1-2a}), (%
\ref{rho-s1-2b}) can be obtained from matrices (\ref{rho-s1-1a}), (\ref%
{rho-s1-1b}) by complex conjugation.

All matrices: $\rho^{\mu}$, $\tilde{\rho}^{\mu}$ discussed above, c.f. Eqs.(%
\ref{rho-s0-1a}), (\ref{rho-s0-1b}), (\ref{rho-s1-1a}), (\ref{rho-s1-1b}), (%
\ref{rho-s1-2a}), (\ref{rho-s1-2b}), fulfill the Tzou commutation relations 
\cite{Tzou57, AO, Beckers95} 
\begin{equation}
\rho^{(\lambda}\rho^{\mu}\rho^{\nu)}=g^{(\lambda\mu}\rho^{\nu)},
\label{Tzou}
\end{equation}
more complicated then (\ref{algebra-b}), where $\left( \lambda\ \mu \
\nu\right) $ is the symmetrizer. There is however no conjugation rule for
matrices $\rho^{\mu}$ and $\tilde{\rho}^{\mu}$, for example there is no such
matrix $S$ that $\tilde{\rho}^{\mu}=S\rho^{\mu}S^{-1}$. We shall see in the
next Section that a conjugation rule (charge conjugation) exists if $%
3\times3 $ matrices $\rho^{\mu}$ are extended to $4\times4$ Dirac matrices $%
\gamma^{\mu}$.

Equations (\ref{const-s1-a}), (\ref{const-s1-b}), (\ref{const-s1-c})
considered together are Lorentz covariant, except from space reflection,
since involve all components of the spinors $\hat{\zeta}_{A\dot{B}}$, $%
\eta_{CD}$ in analogy with spin-$0$ case (the same applies to the set of
equations (\ref{const-s1-2a}), (\ref{const-s1-2b}), (\ref{const-s1-2c})).

\section{Subsolutions of the Dirac equation}

Since the constituent equations (\ref{const-s0-1}), (\ref{const-s0-2}) and (%
\ref{const-s1-a}, \ref{const-s1-b}, \ref{const-s1-c}), (\ref{const-s1-d}, %
\ref{const-s1-e}, \ref{const-s1-f}) seem to be fundamental we shall attempt
in the first place to interpret these equations. We shall first interpret
equations (\ref{const-s0-1}), (\ref{const-s0-2}) and identities (\ref%
{identities0-a}), (\ref{identities0-b}).

Equation (\ref{const-s0-1}) and the identity (\ref{identities0-a}), as well
as equation (\ref{const-s0-2}) and the identity (\ref{identities0-b}) can be
written in form of the Dirac equations:

\begin{equation}
\left( 
\begin{array}{cccc}
0 & 0 & p^{0}+p^{3} & p^{1}-ip^{2} \\ 
0 & 0 & p^{1}+ip^{2} & p^{0}-p^{3} \\ 
p^{0}-p^{3} & -p^{1}+ip^{2} & 0 & 0 \\ 
-p^{1}-ip^{2} & p^{0}+p^{3} & 0 & 0%
\end{array}
\right) \left( 
\begin{array}{c}
\psi^{1\dot{1}} \\ 
\psi^{2\dot{1}} \\ 
\chi \\ 
0%
\end{array}
\right) =m\left( 
\begin{array}{c}
\psi^{1\dot{1}} \\ 
\psi^{2\dot{1}} \\ 
\chi \\ 
0%
\end{array}
\right) ,  \label{A}
\end{equation}%
\begin{equation}
\left( 
\begin{array}{cccc}
0 & 0 & p^{0}-p^{3} & p^{1}+ip^{2} \\ 
0 & 0 & p^{1}-ip^{2} & p^{0}+p^{3} \\ 
p^{0}+p^{3} & -p^{1}-ip^{2} & 0 & 0 \\ 
-p^{1}+ip^{2} & p^{0}-p^{3} & 0 & 0%
\end{array}
\right) \left( 
\begin{array}{c}
\psi^{2\dot{2}} \\ 
\psi^{1\dot{2}} \\ 
\chi \\ 
0%
\end{array}
\right) =m\left( 
\begin{array}{c}
\psi^{2\dot{2}} \\ 
\psi^{1\dot{2}} \\ 
\chi \\ 
0%
\end{array}
\right) ,  \label{B}
\end{equation}
respectively, with one zero component. Equation (\ref{A}) can be written as $%
\gamma^{\mu}p_{\mu}\Psi=m\Psi$ with spinor representation of the Dirac
matrices, $\gamma^{0}=\left( 
\begin{array}{cc}
\mathbf{0} & \sigma^{0} \\ 
\sigma^{0} & \mathbf{0}%
\end{array}
\right) $, $\gamma^{j}=\left( 
\begin{array}{cc}
\mathbf{0} & -\sigma^{j} \\ 
\sigma^{j} & \mathbf{0}%
\end{array}
\right) $, $j=1,2,3$, $\gamma^{5}=\left( 
\begin{array}{cc}
\sigma^{0} & \mathbf{0} \\ 
\mathbf{0} & -\sigma^{0}%
\end{array}
\right) $, $\Psi=\left( \psi^{1\dot{1}},\psi^{2\dot{1}},\chi,0\right) ^{T}$.
Equation (\ref{B}) can be analogously written as $\left(
\gamma^{0}p^{0}-\gamma^{1}p^{1}+\gamma^{2}p^{2}+\gamma^{3}p^{3}\right)
\Phi=m\Phi$, $\Phi=\left( \psi^{2\dot{2}},\psi^{1\dot{2}},\chi,0\right) ^{T}$%
.

We shall demonstrate now that equations (\ref{A}) and (\ref{B}) are charge
conjugated one to another. Complex conjugation of Eq.(\ref{A}) yields:%
\begin{equation}
\left( -1\right) \left( 
\begin{array}{cccc}
0 & 0 & p^{0}+p^{3} & p^{1}+ip^{2} \\ 
0 & 0 & p^{1}-ip^{2} & p^{0}-p^{3} \\ 
p^{0}-p^{3} & -p^{1}-ip^{2} & 0 & 0 \\ 
-p^{1}+ip^{2} & p^{0}+p^{3} & 0 & 0%
\end{array}%
\right) \left( 
\begin{array}{c}
\psi ^{1\dot{1}} \\ 
\psi ^{2\dot{1}} \\ 
\chi \\ 
0%
\end{array}%
\right) ^{\ast }=m\left( 
\begin{array}{c}
\psi ^{1\dot{1}} \\ 
\psi ^{2\dot{1}} \\ 
\chi \\ 
0%
\end{array}%
\right) ^{\ast },  \label{A*}
\end{equation}%
i.e. $\left( -1\right) \left( \gamma ^{0}p^{0}-\gamma ^{1}p^{1}+\gamma
^{2}p^{2}-\gamma ^{3}p^{3}\right) \Psi ^{\ast }=m\Psi ^{\ast }$ where $%
^{\ast }$ denotes complex conjugation. Acting from the left with matrix $%
\gamma ^{3}$ on Eq.(\ref{A*}) we obtain equation $\left( \gamma
^{0}p^{0}-\gamma ^{1}p^{1}+\gamma ^{2}p^{2}+\gamma ^{3}p^{3}\right) \gamma
^{3}\Psi ^{\ast }=m\gamma ^{3}\Psi ^{\ast }$ which has the same form as Eq.(%
\ref{B}) (the charge conjugation matrix $C$ is thus defined as $C\gamma
^{0}\equiv \gamma ^{3}$ \cite{BD64}). Hence the initial equations (\ref%
{const-s0-1}), (\ref{const-s0-2})) are charge conjugated one to another in a
sense that they are charge conjugated after extension to the Dirac form.

Similar considerations lead to conclusion that also Eqs.(\ref{const-s1-a}, %
\ref{const-s1-b}, \ref{const-s1-c}), (\ref{const-s1-d}, \ref{const-s1-e}, %
\ref{const-s1-f}) and identities (\ref{identities1-a}), (\ref{identities1-b}%
) as well as Eqs.(\ref{const-s1-2a}), (\ref{const-s1-2b}) and identities (%
\ref{identities1-c}), (\ref{identities1-d}) can be written in Dirac form to
reveal that they are charge conjugated one to another.

The observations made above can be given representation independent
formulation. Let us notice that the three component equations, for instance (%
\ref{const-s0-1}), (\ref{const-s0-2}) as well as the identities (\ref%
{identities0-a}), (\ref{identities0-b}), can be obtained by projecting the
Dirac equation with projection operator $P_{4}=$\textrm{diag}$\left(
1,1,1,0\right) $. Incidentally, there are other projection operators which
lead to analogous three component equations, $P_{1}=$\textrm{diag}$\left(
0,1,1,1\right) $, $P_{2}=$\textrm{diag}$\left( 1,0,1,1\right) $, $P_{3}=$%
\textrm{diag}$\left( 1,1,0,1\right) $ but we shall need only the operator $%
P_{4}$.

In general, we can consider subsolutions, of form $P_{4}\Psi $, of the Dirac
equation: 
\begin{equation}
\gamma ^{\mu }p_{\mu }P_{4}\Psi =mP_{4}\Psi ,  \label{D1}
\end{equation}%
which is equivalent to (\ref{A}) in the case of spinor representation of the
Dirac matrices.

Accordingly, acting from the left on (\ref{D1}) with $P_{4}$ and $\left(
1-P_{4}\right) $ we obtain two equations: 
\begin{subequations}
\begin{align}
P_{4}\left( \gamma ^{\mu }p_{\mu }\right) P_{4}\Psi & =mP_{4}\Psi ,
\label{D2a} \\
\left( 1-P_{4}\right) \left( \gamma ^{\mu }p_{\mu }\right) P_{4}\Psi & =0.
\label{D2b}
\end{align}%
In the spinor representation of $\gamma ^{\mu }$ matrices Eq.(\ref{D2a}) is
equivalent to (\ref{const-s0-1}), while (\ref{D2b}) is equivalent to the
identity (\ref{identities0-a}).

Now the projection operator can be written as $P_{4}=\frac{1}{4}\left( 3%
\mathbf{+}\gamma ^{5}-\gamma ^{0}\gamma ^{3}+i\gamma ^{1}\gamma ^{2}\right) $
(and similar formulae can be given for other projection operators $%
P_{1},P_{2},P_{3}$), i.e. all equations (\ref{D1}), (\ref{D2a}), (\ref{D2b})
are now given representation independent form. Let us also note that the
projection operator $P_{4}$ commutes with two generators of Lorentz
transformations $\gamma ^{0}\gamma ^{3}$ and $\gamma ^{1}\gamma ^{2}$ (and
does not commute with other generators), i.e. is invariant under boosts in $%
x^{0}x^{3}$ plane and rotations in $x^{1}x^{2}$ plane. Accordingly, the
three component equations are covariant with respect to such Lorentz
transformations only. Let us note finally that all three component equations
describe particles with definite mass and only one component of spin
defined. Results of this Section are directly generalized for the case of
interaction introduced via minimal coupling, $p^{\mu }\rightarrow \pi ^{\mu
}=p^{\mu }-eA^{\mu }$.

\section{Discussion}

We have shown that spinor formalism discloses internal structure of KDP
equations which manifests itself by presence of redundant components - there
are special three-component solutions of these equations. Accordingly, the
meson spin-$0$ and spin-$1$ KDP equations split into pairs of
three-component constituent equations, each equation describing a particle
with definite mass and only one component of spin defined (all
three-component constituent equations discussed above are similar in a sense
that their matrices $\rho ^{\mu }$ fulfill the same commutation relations (%
\ref{Tzou}) \cite{Tzou57, AO, Beckers95}) and in the case of spin-$1$ KDP
equations an additional wavefunction fulfilling the Klein-Gordon equation is
present. Moreover, solutions of the constituent equations are subsolutions
of appropriate Dirac equations and pairs of such Dirac equations,
corresponding to pairs of constituent equations, are charge conjugated one
to another. This last finding entitles us to conclude that
Kemmer-Duffin-Petiau equations describe mesons as composed from
quark-antiquark pairs. These results are consistent with quark theory of
mesons \cite{Gell-Mann}. Furthermore, existence of additional constituent
particle with the same mass in the case of spin-$1$ mesons elucidates
problem of meson multiplets. It is well known that pseudoscalar mesons can
be approximately arranged in $SU(3)$ octets while in the case of vector
mesons octets are strongly mixed with $SU(3)$ singlets to form nonets \cite%
{Sakurai}. Since we get two constituent, charge conjugated equations, plus a
single equation it might be inferred that this picture is consistent with
octet-singlet mixing if we interpret constituents in three-component
equations as quarks and an additional constituent as a $SU(3)$ singlet
vector meson. The mixing is indeed present since wavefunction fulfilling (%
\ref{const-s1-g}) is dynamically coupled to wavefunctions in three-component
equations (\ref{const-s1-a}, \ref{const-s1-b}, \ref{const-s1-c}), (\ref%
{const-s1-d}, \ref{const-s1-e}, \ref{const-s1-f}).

Let us stress that the separation of a meson into constituents is imperfect,
since although each of the constituent equations describes a massive
particle, its Lorentz symmetry is broken. This offers explanation of quark
confinement different than in quantum chromodynamics where linearly
increasing potential energy between a quark and other quarks in a hadron is
responsible for confinement \cite{Takahashi}). We hope that our results can
also cast light on the problem of spin crisis \cite{Ma}\ since meson
constituents in our theory have one component of spins defined only. Let us
assume that proton constituents (quarks) have the same nature as the meson
constituents of our theory. It follows that the proton spin cannot be
obtained as a sum of spins of its constituents since the constituents have
not fully defined spins.

\begin{acknowledgement}
The author expresses his gratitude to Prof. K. Zalewski for discussion.
\end{acknowledgement}

\end{subequations}

\end{document}